\documentclass[amsmath,amssymb,amsfont,aps,prl,groupedaddress,twocolumn,notitlepage, nobibnotes,nofootinbib,longbib]{revtex4-2}
 \pdfoutput=1

\usepackage[utf8]{inputenc}
\usepackage[english]{babel}
\usepackage{physics}
\usepackage{graphicx}
\usepackage{multirow}
\usepackage{amsmath,bm}
\usepackage{booktabs}
\usepackage{url}
\usepackage{soul}
\usepackage{array}
\usepackage[colorlinks,citecolor=blue,linkcolor=blue,urlcolor=blue, breaklinks=true]{hyperref}
\usepackage{float}

\usepackage{xcolor}
\usepackage{xspace}
\usepackage{ifthen}

      \newcommand{\showcomments}{true}

\newcommand{\andrea}[1]
{\ifthenelse{\equal{\showcomments}{true}}{{\color{orange} \textbf{Andrea says: }#1}}{\xspace}}

\newcommand{\carlo}[1]
{\ifthenelse{\equal{\showcomments}{true}}{{\color{violet} \textbf{Carlo says: }#1}}{\xspace}}

\newcommand{\marios}[1]%
{\ifthenelse{\equal{\showcomments}{true}}%
{{\color{blue}{\small \textbf{Marios says:} #1}}}{\xspace}}%

\newcommand{\emanuele}[1]%
{\ifthenelse{\equal{\showcomments}{true}}%
{{\color{red}{\small \textbf{Emanuele says:} #1}}}{\xspace}}%

\begin{document}

\title{Photonic Implementation of Quantum Gravity Simulator}

\date{\today}
\author{Emanuele Polino}
\author{Beatrice Polacchi}
\author{Davide Poderini}
\author{Iris Agresti}
\author{Gonzalo Carvacho}
\author{Fabio Sciarrino}
\email{fabio.sciarrino@uniroma1.it}
\affiliation{Dipartimento di Fisica, Sapienza Universit\`{a} di Roma, P.le Aldo Moro 5, I-00185 Roma, Italy}

\author{Andrea \surname{Di Biagio}}
\affiliation{Dipartimento di Fisica, Sapienza Universit\`a di Roma, Piazzale Aldo Moro 5, Roma, Italy}
\affiliation{Institute for Quantum Optics and Quantum Information (IQOQI) Vienna, Austrian Academy of Sciences, Boltzmanngasse 3, A-1090 Vienna, Austria}

\author{Carlo Rovelli}
\email{rovelli@cpt.univ-mrs.fr}
\affiliation{Aix-Marseille University, Universit\'e de Toulon, CPT-CNRS, Marseille, France,}
\affiliation{Department of Philosophy and the Rotman Institute of Philosophy, Western University, London ON, Canada,}
\affiliation{Perimeter Institute, 31 Caroline Street N, Waterloo ON, Canada}

\author{Marios Christodoulou}
\email{marios.christodoulou@oeaw.ac.at}
\affiliation{Institute for Quantum Optics and Quantum Information (IQOQI) Vienna, Austrian Academy of Sciences, Boltzmanngasse 3, A-1090 Vienna, Austria}
\affiliation{Vienna Center for Quantum Science and Technology (VCQ), Faculty of Physics, University of Vienna, Boltzmanngasse 5, A-1090 Vienna, Austria}

\begin{abstract}

Detecting gravity mediated entanglement can provide evidence that the gravitational field obeys quantum mechanics. We report the result of a simulation of the phenomenon using a photonic platform. The simulation tests the idea of probing the quantum nature of a variable by using it to mediate entanglement, and yields theoretical and experimental insights.
We employed three methods to test the presence of entanglement: Bell test, entanglement witness and quantum state tomography. We also simulate the alternative scenario predicted by gravitational collapse models or due to imperfections in the experimental setup and use quantum state tomography to certify the absence of entanglement. Two main lessons arise from the simulation: 1) which--path information must be first encoded and subsequently coherently erased from the gravitational field, 2) performing a Bell test leads to stronger conclusions, certifying the existence of gravity mediated nonlocality.
\end{abstract}

\maketitle
\raggedbottom

The gravitational field is generally expected to obey quantum mechanics, like any other physical field. But to this day there is no experimental evidence that this is the case.  At the 1957 Chapel Hill conference, Richard Feynman famously emphasized that the gravitational field can be set into quantum superposition by simply setting a source, namely a mass, into the superposition of two positions \cite{dewitt-morette2011role}.  But, given the weakness of the gravitational interaction, how can we find empirical evidence for the superposition of field configurations?

The past few years have seen an intense interest in the possibility of obtaining such evidence on the laboratory bench, by detecting  entanglement generated between quantum masses interacting gravitationally \cite{marletto2017gravitationallyinduced,bose2017spin,krisnanda2020observable}. The key idea is that if the field mediating an interaction can be in a quantum superposition, 
this interaction can entangle degrees of freedom.
A well known result in quantum information theory  states that quantum entanglement cannot be created between two systems by local operations and classical communication (LOCC) \cite{bennett1996concentrating,bennett1996mixed,popescu1997thermodynamics}.  Gravity Mediated Entanglement (GME) can thus provide evidence that the gravitational field that mediates the interaction is in a quantum superposition \cite{christodoulou2019possibility}. Rapid advances in quantum control of larger masses \cite{delic2020cooling,magrini2021real,tebbenjohanns2021quantum,margalit2021realization} and measuring the gravitational field of smaller masses \cite{westphal2021measurement,barzanjeh2022optomechanics}, might soon make this momentous experimental test possible. 

The intriguing possibility of conducting `table top' quantum gravity experiments has attracted such interest that it is rapidly developing into a new field, exploring different possible experimental setups and debating their theoretical implications.


The effect is predicted by most current tentative quan-
tum gravity theories, such as loop quantum gravity,
string theory, as well as low energy effective field theory. On the contrary, 
it is not predicted by theories where the gravitational interaction is mediated by a local classical field \cite{oppenheim2018postquantum,pearle1989combining,percival1995quantum}, nor by theories where the gravitational field does not display sufficiently long lasting massive superpositions of macroscopically different configurations or where massive superpositions spontaneously collapse \cite{penrose1996gravity,diosi1987universal,bassi2013models,carney2019tabletop}. Thus, a negative result of the experiment would also be of particular interest, as the absence of GME would falsify common assumptions in quantum gravity research and provide evidence for  these unorthodox theories.

The logic of the GME experiments, however, is more subtle:  find evidence for a key property of certain collective degrees of freedom, by looking at the way these allow other degrees of freedom to get entangled. Unpacking this argument is the subject of a lively debate on the precise epistemological conclusions that can be drawn from the detection of GME. Recent overviews of the debate are given in \cite{fragkos2022inference, huggett2022quantum}. 

In this work, we report the photonic implementation of a quantum circuit simulating the experimental proposal of \cite{bose2017spin}. The simulation sheds light on subtle aspects of the logic behind the claim that detecting GME is tantamount to evidence for the quantum nature of the gravitational interaction. Using an entanglement witness and the violation of a Bell inequality, we study how these different measurement protocols can certify the presence of entanglement given realistic levels of noise. To study the possibility of a negative experimental outcome, we simulate spontaneous collapse models by introducing decoherence into the simulator and employ quantum state tomography to certify the absence of entanglement.

\begin{figure}[h!]
    \includegraphics[width=1\columnwidth]{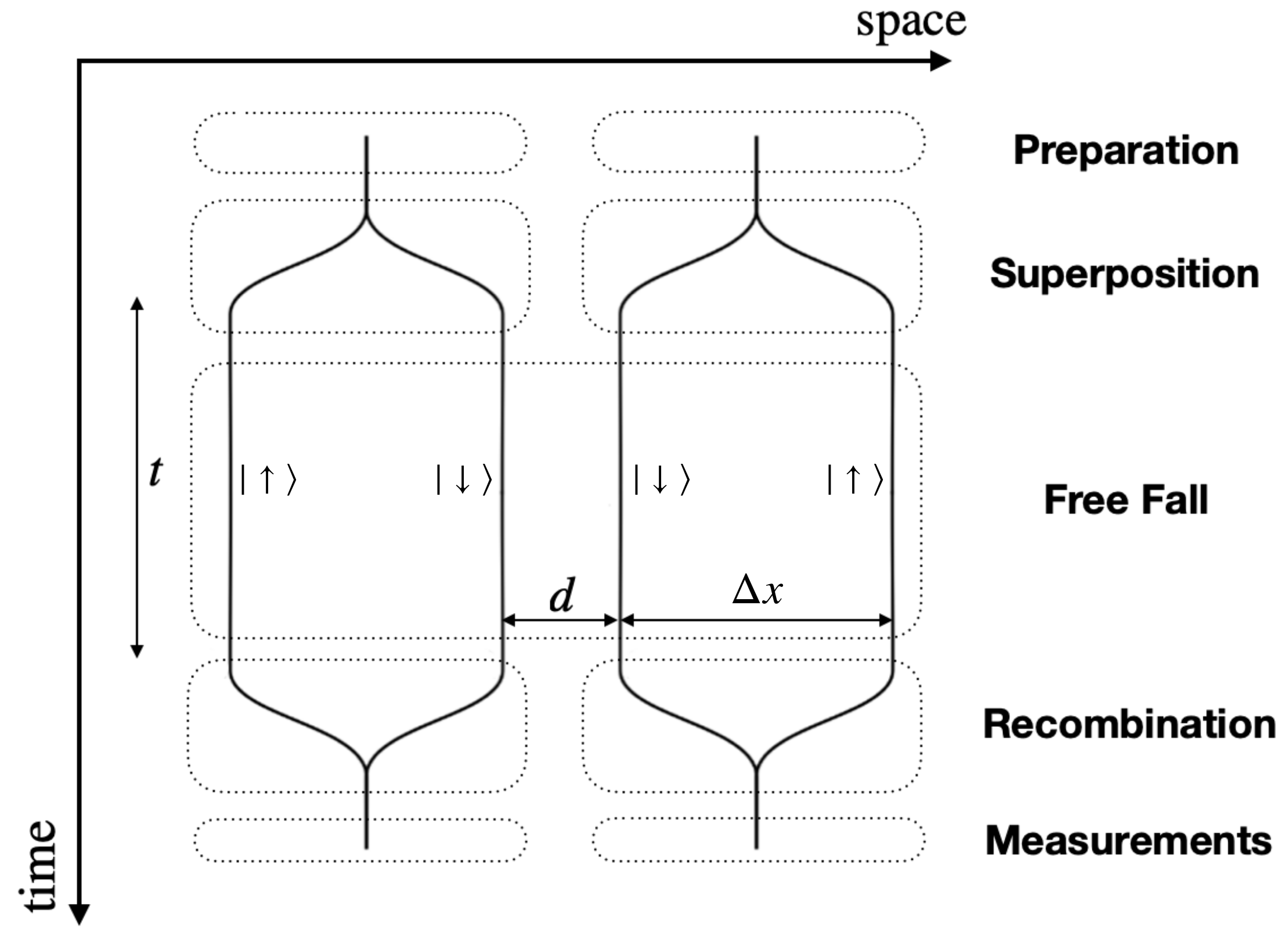}
   \caption{\textbf{Two masses in path superposition interacting gravitationally become entangled.} Two massive particles with embedded magnetic spin are put into a spin-dependent path superposition. They are then left to Free Fall, where they interact via the gravitational field only. Then, the path superposition is undone and measurements are performed on the spins. During the Free Fall, each branch of the superposition accumulates a different phase, which entangles the two particles.}
\label{fig:GME}
\end{figure}

In the experiment, two masses are manipulated into a macroscopic center of mass superposition through an inhomogeneous magnetic field that couples to a spin (NV-center) embedded in each mass, see Fig.~\ref{fig:GME}. 
Once the superposition is accomplished, the state of the full quantum system is
\begin{equation}\label{eq:GME_superposition}
    \ket{\uparrow\uparrow}\ket{g_{LR}}+\ket{\uparrow\downarrow}\ket{g_{LL}}
    +\ket{\downarrow\uparrow}\ket{g_{RR}}+\ket{\downarrow\downarrow}\ket{g_{RL}},
\end{equation}
where $\ket\uparrow$ and $\ket\downarrow$ are spin-$z$ eigenstates, and the states $\ket{g_{XY}}$ are coherent states of the gravitational field and centre of mass position of the masses \cite{christodoulou2019possibility,christodoulou2022locally}. These states are approximate energy eigenstates \cite{chevalier2020witnessing}, they simply accumulate a relative phase during the Free Fall stage.
Once the superposition is undone, the spins disentangle from the geometry. Tracing it out, we get
\begin{equation}\label{eq:GME_spins}
    e^{i\phi_{LR}}\ket{\uparrow\uparrow}+e^{i\phi_{LL}}\ket{\uparrow\downarrow}
    +e^{i\phi_{RR}}\ket{\downarrow\uparrow}+e^{i\phi_{RL}}\ket{\downarrow\downarrow}.
\end{equation}
For generic values of the phases, this is an entangled state.  A detailed covariant derivation of this effect is given in \cite{christodoulou2022locally}.

\begin{figure*}[ht!]
    \centering
    \includegraphics[width=0.8\textwidth]{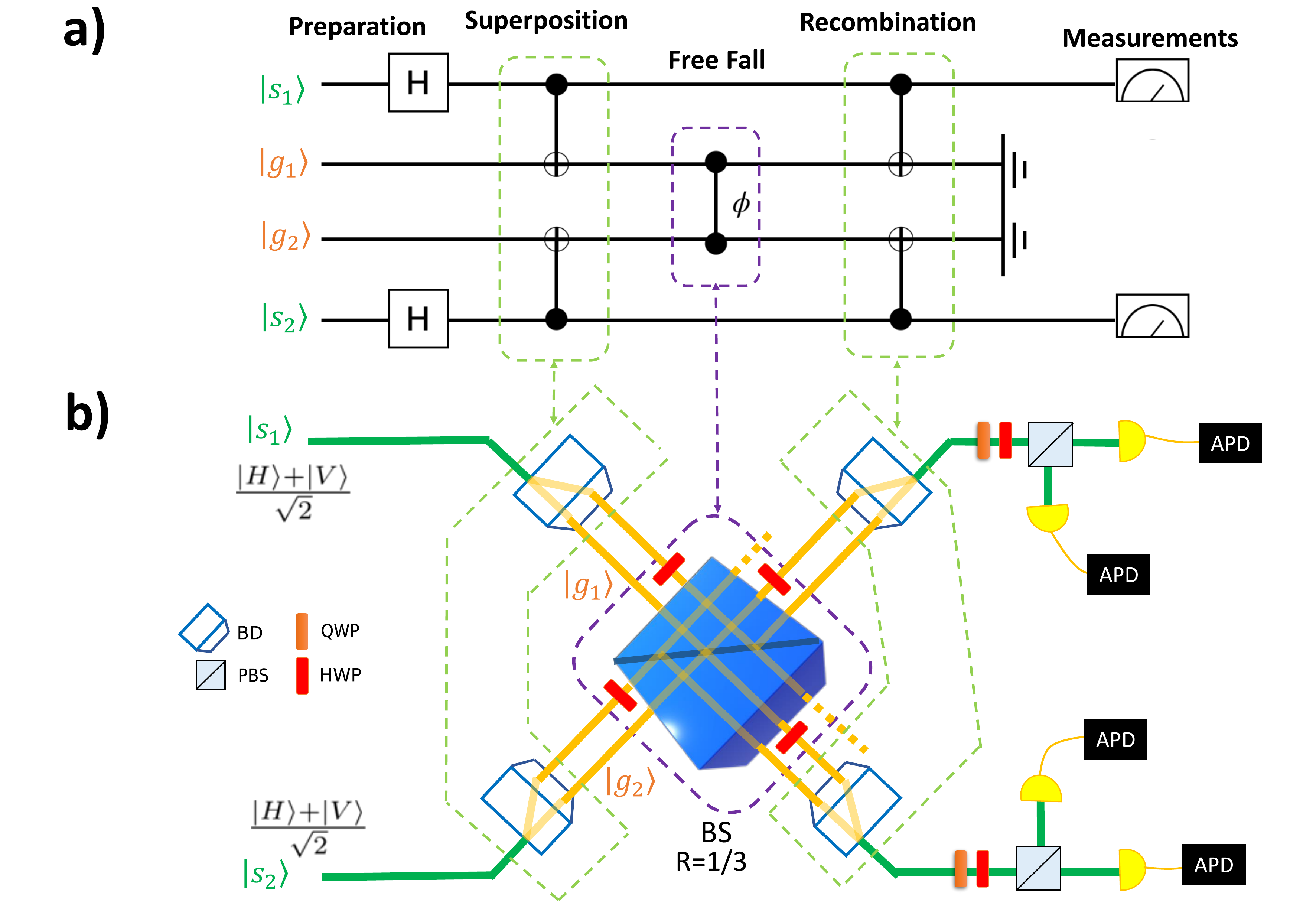}
    \caption{\textbf{The quantum circuit simulator (a) and its photonic implementation (b).}    
    \textbf{a)} Two qubits, $\ket{s_1}$ and $\ket{s_2}$, represent the spin degrees of freedom, while two qubits,  $\ket{g_1}$ and $\ket{g_2}$, represent the geometry. Each stage of the experiment is mapped into quantum gates acting on the qubits. 
        \textbf{b)} The simulator is implemented by using the path and polarization degrees of freedom of two photons. The spin qubits of the simulator are encoded in the polarization degree of freedom of the photons, while the geometry degrees of freedom are encoded in the photon paths. The two photons are independently prepared in a superposition of horizontal and vertical polarization and each one passes through a beam displacer (BD), which completely entangles the path of each photon with its polarization. The CZ gate is implemented thanks to bosonic interference, due to the indistinguishability of the photons, at the beam splitter (BS). Two half-waveplates (HWP) momentarily make the polarization of all paths equal in order to allow the realization of the Control-Phase (CZ) gate on this degree of freedom. Finally, the qubit state is restored by two other half-waveplates and the paths are recombined by final BDs, which disentangles path and polarization. Finally, the polarizations of the photons are measured by means of quarter- and half-waveplates (QWP and HWP)and polarizing beam splitter (PBS) followed by single photon detectors (APDs).  
    }
    \label{fig:Simulators} 
\end{figure*}

The quantum circuit simulator is depicted in Fig.~\ref{fig:Simulators}a\footnote{The quantum circuit of Fig.~\ref{fig:Simulators}a has been also discussed in \cite{bhole2020witnesses}.}. It is a straightforward representation of the evolution of the experiment in the regime $\Delta x\gg d$, where only the phase in the branch of closest approach needs to be considered. This regime simplifies the analysis, without compromising the physics \cite{bose2017spin,christodoulou2019possibility}.

The circuit represents the two spins and the geometry as a $16$-dimensional system.
Each embedded spin is simulated with a qubit, while the geometry degrees of freedom with two qubits (a ququart). 
We write vectors as belonging to the Hilbert space
\begin{equation}
    \mathbb{C}^2\otimes\mathbb{C}^4\otimes\mathbb{C}^2 = \mathcal{H}_{\mathrm{spin}_A}\otimes\mathcal{H}_{\mathrm{geometry}}\otimes\mathcal{H}_{\mathrm{spin}_B}.
\end{equation}
At the end of the Free Fall stage, the state of the full system is
\begin{equation}\label{eq:QC_freefall}
 \frac{1}{2}\big(\ket{0000}+\ket{0011}+\ket{1100}+e^{i\phi}\ket{1111}\big).
\end{equation}
After the Recombination stage, the state of the geometry ququart factorises. At the moment of measurement, the spin qubits are in the state
\begin{equation}\label{eq:QC_final_spins}
    \frac12\big(\ket{00}+\ket{01}+\ket{10}+e^{i\phi}\ket{11}\big).
\end{equation}

We implemented the quantum circuit simulator on the photonic platform shown in Fig.~\ref{fig:Simulators}b. The polarization of the photons carries the qubits representing the spins, while the geometry ququart is encoded in the paths of the photons.
Photons of wavelength $\sim785\,\mathrm{nm}$ are produced by Spontaneous Parametric Down Conversion (SPDC) from a nonlinear barium borate (BBO) crystal pumped by a pulsed laser at $392.5\,\mathrm{nm}$.
The CNOT gates of the Superposition and Recombination stages are deterministically performed on each photon's path-polarization space through calcite beam displacers (BD).  The Control-Phase gate with a phase equal to $\pi$ acting on the paths of the photon, which represents the effect of the Free Fall stage, is realized by a probabilistic scheme exploiting bosonic interference in a beam splitter with reflection coefficient $1/3$  \cite{obrien2003demonstration}.  
Further details on the scheme and the experimental setup are provided in the Supplementary Material.

At the end of the Recombination stage, the polarization state of the photons is
\begin{equation}\label{eq:final_state_QC}
    \frac12\big(\ket{VV}+\ket{HV}+\ket{VH}-\ket{HH}\big)\;,
\end{equation}
where horizontal ($\ket{H}$) and vertical ($\ket{V}$) polarizations encode respectively the qubit states $\ket{1}$ and $\ket{0}$ in Eq. \eqref{eq:QC_final_spins}.

Before the final measurement, we apply the local unitary operation $(\sigma_z + \sigma_x)/\sqrt{2}$ on the second qubit, by means of a half-wave plate rotated by $22.5^{\circ}$ with respect to its optical axis. This step allows to recast the state as the maximally entangled singlet state
\begin{equation}
    \ket{\Psi^-} = \frac{1}{\sqrt{2}}\big(\ket{HV}-\ket{VH}\big).
\label{eq:final_state_singlet}
\end{equation}
This final rotation, which is equivalent to changing the measurement basis, simplifies the analysis without loss of generality.\\


\begin{figure*}[ht!]
    \centering
    \includegraphics[width=.8\textwidth]{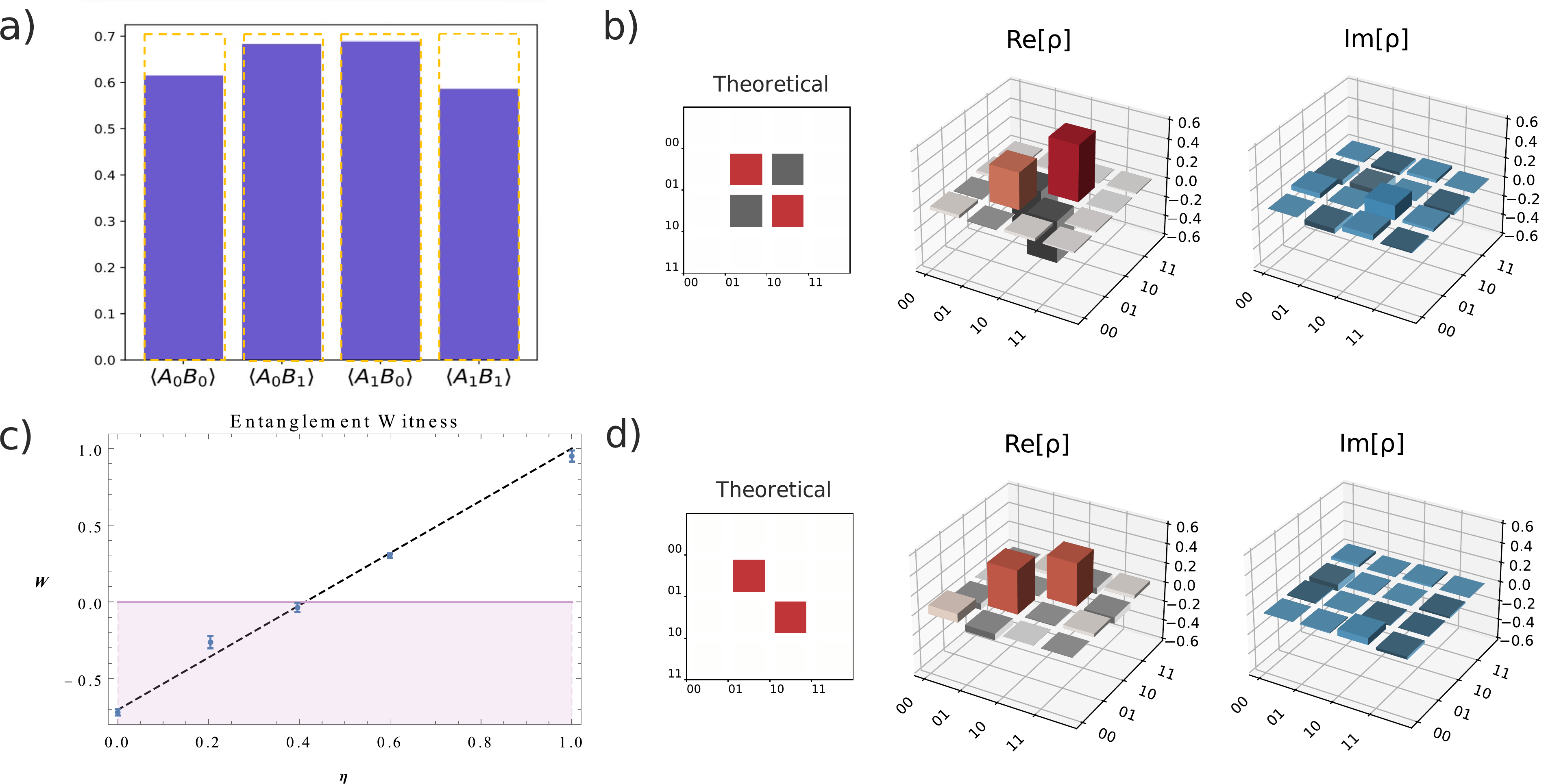}
    \caption{\textbf{Results of the simulator, without and with decoherence.} 
    {\bf a)} Expectation values of the operators used for CHSH test on the spin qubits. The lighter colored parts in each bar  (hardly visible) represent the Poissonian experimental errors associated to each observable. The orange dashed bars are the values expected from an ideal maximally entangled state.
     {\bf b)} Real and imaginary parts of the measured density matrix of the spin qubits.
    {\bf c)} Measured values of the entanglement witness $\mathcal{W}$  as function of the degree of decoherence $\eta$. The latter corresponds to the relative time delay of different polarization normalized to the coherence time of the photons. The purple shaded area indicates the region where the witness certifies the entanglement of the state.  The dashed black line represents the theoretical curve from the model of the experimental setup. Error bars are due to Poissonian statistics of the measured events.  {\bf d)} Real and imaginary parts of the measured density matrix of the polarization state of the spin qubits, where the state has experienced maximum  decoherence effects ($\eta=1$) introduced by a delay between linear polarizations greater than the photon coherence time. The off-diagonal terms are completely suppressed, and the state is separable.}
\label{fig:expresQC}
\end{figure*}

To certify the presence or absence of entanglement at the Measurement stage we implemented three strategies: quantum state tomography, an entanglement witness, and the violation of a Bell inequality. Each comes with its own merits and shortcomings, which we briefly recall.

Quantum state tomography \cite{james2001measurement} provides the maximum amount of information about a quantum system, by measuring enough observables to fully reconstruct the quantum state. Of the three methods considered, this is the only one capable of certifying the \textit{absence} of entanglement.  
Quantum state tomography requires the implementation of a large number of measurements that for systems of larger dimensions can be too expensive to be performed.
Entanglement can be detected with fewer resources by means of entanglement witnesses \cite{horodecki2009quantum,guhne2009entanglement,friis2019entanglement}, that are observables $W$ such that $\langle W\rangle\geq 0$ for all separable states and $\langle W\rangle<0$ for at least one entangled state. Therefore, a negative expectation value implies the state is entangled. 
Alternatively, the violation of a Bell inequality, such as the CHSH inequality \cite{clauser1969proposed}, allows to draw conclusions with strictly weaker assumptions than entanglement witnesses  and tomography \cite{bell1964einstein,brunner2014bell}. This is because, in contrast to the two previous techniques, Bell inequalities do not rely on assuming the validity of quantum theory nor the correct implementation of the quantum measurements, that is, it provides what is commonly called a \emph{device independent} certification of the presence of entanglement, one that relies on minimal assumptions about the experimental setup. Note, however, that not all entangled states can violate a Bell inequality, while an entanglement witness can be designed to detect arbitrary amounts of entanglement.

After the fine alignment of the setup, we performed a CHSH test on the polarization of the photons at the end of the circuit, obtaining a value $S^{\text{exp}} = 2.402 \pm 0.015$. That is, the classical bound of $2$ was violated by more than $26$ standard deviations. This provides a device independent certification that entanglement was successfully mediated via the action of the CZ gate on the path degree of freedom of the photons. Then, we measured the following entanglement witness
\begin{equation}
    \mathcal{W} = 1-| \langle\sigma_x\otimes\sigma_x\rangle+\langle\sigma_y\otimes\sigma_y\rangle|,
    \label{eq:witnesss}
\end{equation}
obtaining a value of $\mathcal{W}^{\text{exp}}= -0.72 \pm 0.02 $. This result violates the separable bound of $0$ by more than $36$ standard deviations, where the statistical uncertainty has been computed  assuming Poissonian statistics. The entanglement witness \eqref{eq:witnesss} is equivalent, up to local unitaries,  to the one proposed in Ref.~\cite{bose2017spin}. Results from the quantum state tomography of the generated state are given in Fig.~\ref{fig:expresQC}b.  





To simulate the effects of spontaneous collapse, we induced decoherence by the implementation of time-delays across different photon polarizations at the output of the beam splitter. As long as the delay is shorter than the coherence time of the photon wavepacket, some degree of entanglement can be generated and detected by the witness. However, when the delay is longer than the photon coherence time, no entanglement can be detected in the final state.

If one of the photons passes through a birefringent slice after the beam splitter, the state before measurement becomes 
\begin{equation} \label{eq:timeent}
\ket{\Psi}_\mathrm{del} = \frac{1}{\sqrt{2}}(\ket{H}\ket{V}_{t_V}-\ket{V}\ket{H}_{t_H}).
\end{equation}
Here, $t_H$ and $t_V$ are the distinguishable delays acquired respectively by the horizontal and vertical polarizations.  
When the delay is greater than the coherence time of the photons, tracing out the information regarding the delays results in the completely mixed state 
\begin{equation} \label{eq:mixedpol}
\rho_\mathrm{mix}^\mathrm{pol} = \frac12\big(\ketbra{HV}{HV}+\ketbra{VH}{VH}).
\end{equation}
In contrast, when the delay is shorter than the coherence time of the photons, the state is 
\begin{equation} 
(1-\eta) \ketbra{\Psi^-}{\Psi^-}+ \eta\; \rho_\mathrm{mix}^\mathrm{pol},
\end{equation}
which is partially mixed.
The parameter $\eta$ quantifies the amount of decoherence due to the polarization-dependent delay.
Changing the thickness of the birefringent slices allows to vary $\eta$ from vanishing to maximum decoherence.


We measured the witness $\mathcal{W}$ and performed state tomography for five values of $\eta$.  For $\eta>0.4$, which correspond to delays greater than around $450\,\mathrm{ps}$, the observed witness does not violate the separable bound, see Fig. \ref{fig:expresQC}c. The state tomography for the completely decohered state ($\eta=1$) is reported in Fig. \ref{fig:expresQC}d. Results for all values of $\eta$ are reported in Fig. 2 of the Supplementary Information. 

Quantum state tomography allows to exclude the presence of entanglement in a quantum mechanical framework. Indeed, failure to detect entanglement with the witness is not proof of the absence of entanglement. For example, the entanglement witness did not detect entanglement for $\eta\sim0.6$, but a positive partial transpose (PPT) test \cite{horodecki2009quantum} on the results of state tomography revealed the presence of entanglement. The four eigenvalues were\footnote{The error intervals were computed with the Monte Carlo method.} $(0.583, 0.357, 0.230,-0.170)\pm (0.005,0.007,0.007,0.008)$. Note that, for two-qubit states, the presence of a negative eigenvalue in the partial transpose is a necessary and sufficient condition for entanglement.

We also simulated a conceptually different effect by introducing noise in the state of the geometry ququart. Indeed, while decoherence effects could be caused by spontaneous collapse of the state of the particles, in a realistic experiment it may also be the case that the interaction among particles is not strong enough to generate observable effects. For example, this would be the case if the distance between the interferometers paths is too large, or if the particles pass through the interferometers at different times.
Therefore, the expected partially distinguishable state $\rho_\mathrm{part.~dist.}$ will be a mixture of the following form:
\begin{equation}
    \rho_\mathrm{part.~dist.} = v \ketbra{\Psi^-}{\Psi^-} + (1-v) \rho_\mathrm{dist} \; ,
\label{eq:HOM_state}
\end{equation}
where the visibility $v$ depends on the time delay between the photons. The density matrix $\rho_\mathrm{dist.}$ is the expected two-photon state when the delay time is longer than the coherence time of the photons and is given by
\begin{equation}
\begin{split}
    \rho_\mathrm{dist} &= \frac12 \big( \ketbra{H+}{H+} +\ketbra{+H}{+H} \big),
\end{split}
\label{eq:rho_dist}
\end{equation}
with $\ket{+} = ( \ket{H} + \ket{V} )/\sqrt{2}$. 
We implemented this by varying the relative time arrival of the photons in the probabilistic control--gate. The measured entanglement witnesses $\mathcal{W}$ and quantum state tomographies for different indistinguishability degrees $v$ are reported in Fig. 4 and Fig. 3 of the Supplementary Information. The results of the tomography for the two kinds of decoherence models are qualitatively different. \\
 
\emph{Conclusions.} We detected the creation of mediated entanglement in the photonic simulator using three different methods: Bell inequality violation, entanglement witness, and quantum state tomography. We simulated two kinds of decoherence, due to noise and due to new physics, and noted they may be distinguished with state tomography.






The study of the GME experimental proposals is merging the scientific culture of the two research communities of quantum gravity and quantum information. It is shedding new light theoretically and experimentally on the possibilities of quantum gravity phenomenology and has sparked a lively debate which reaches to the foundations of quantum theory.

A core aspect of the debate is the concept of witnessing the non-classicality of the gravitational field, and more in general the distinction between `classical' and `non-classical' behaviour. Within quantum mechanics, superposition is a hallmark of non-classical behaviour. Superposition is, however, a theory- and basis-dependent concept.
Its operational content can be encapsulated in the existence of non-commuting observables. The existence of non-commuting observables is the defining characteristic of non-classical systems employed in the recent theory-independent generalisations of the LOCC theorems used to analyse GME \cite{marletto2020witnessing,galley2021nogoa}.

{Our simulation makes explicit the crucial role of the non-commuting variables of the gravitational mediator. The mediating ququart entangles with the spin qubits due to the presence of the $\mathrm{X}\otimes \mathrm{I}$ and $\mathrm I \otimes \mathrm{X}$ observables in the $\mathrm{CNOT}$ gates. The phases are then generated by the $\mathrm{ZZ}$ observable. Introducing decoherence in the mediator removes the non-commutativity since $\tr\mathrm{XY I} = \tr \mathrm{YX I}$. Once the non-commutativity is removed, no entanglement is generated between the spins.}



While high level abstract tools like the LOCC theorems can be very powerful, in practice, they are only informative insofar as they can be applied on well developed theories that physicists are interested in testing. Linearized quantum gravity is the effective quantum field theory expected to correctly describe the physics in this regime. The main question within this theory is: what type of excitations of the field are responsible for the mediation of entanglement?

Two proposed answers, the Newtonian potential \cite{anastopoulos2018comment,anastopoulos2021gravitational} and off-shell gravitons \cite{marshman2020locality,bose2022mechanism}, are not satisfying because they are gauge-dependent. A possible gauge-independent answer is on-shell gravitons. Indeed, the mediation of entanglement was shown to be concomitant with the presence of radiation \cite{danielson2021gravitationally,hidaka2022complementarity} in a related thought experiment \cite{mari2016experimentsa,belenchia2018quantum}. Nonetheless, radiation cannot be responsible for entanglement in the GME experiments. If radiation is emitted during the experiment, it will carry away which--path information about the masses and lead to decoherence. Since the objective of the GME experiments is to detect the entanglement by measuring the masses, radiation cannot be involved in the experiment. A partial model of how this takes place is given in \cite{christodoulou2019possibility,christodoulou2022locally}, where the field is in a superposition of semiclassical geometries during the Free Fall stage  and returns to one semiclassical state at the Measurememt stage.

So, what is the mediator carrying quantum information and giving rise to gravity mediated entanglement? In some sense, the entanglement has to be mediated via an interaction that is non-radiative. A hint on what is going on is provided by our simulation. During each run of the simulation, the geometry ququart acquires which-path information about the spin qubits only for a limited amount of time. The initial CNOT gates write the state of the spins in the geometry, while the final CNOT gates erase this information in a coherent way. This is a crucial point: since the geometry ququart is not measured, it would be impossible to detect the entanglement in the spins qubits if they were still entangled with it at the moment of measurement. In the gravitational experiment, \textit{the states of the geometry should temporarily acquire which-path information about the masses, but they should not propagate that information to infinity}. Instead the which--path information must be coherently erased from the field states by the time of measurement.

A second important lesson of the simulation is that, rather than certifying an entanglement witness, \emph{Bell tests would provide more convincing evidence that a quantum gravitational effect has been observed}. Similarly, concluding a definitive negative experimental outcome would require state tomography. Much of the literature on possible experimental protocols for detecting gravity mediated entanglement focuses somewhat misleadingly on employing an entanglement witness to certify, or fail to certify, gravity mediated entanglement. 


Consider first the case of the negative experimental outcome, in which no GME is detected. This measurement would be of extraordinary consequence for fundamental physics, upending convictions held by generations of theoretical physicists. It would immediately falsify all mainstream quantum gravity theories such as loop quantum gravity and string theory, as well as any other approach that claims linearised quantum gravity as its low energy limit. Such a conclusion would be accepted by the wider competent community only once extraordinary evidence is provided. Even after sources of noise have been excluded, it is likely that scientific consensus would form not with the failure to verify an entanglement witness but when high precision state tomography has been performed.  State tomography yields maximal information of the quantum state of the spins and, unlike a Bell test or an entanglement witness, can certify the complete \emph{absence} of entanglement.\footnote{ If precise knowledge of the apparatus cannot be assumed, an alternative method to detect entanglement may be provided by automated optimizations for fully black-box approaches  \cite{poderini2022ab}.}

Consider next the case of the positive outcome. The detection of GME would provide empirical evidence for the existence of quantum gravity, as it would verify a prediction of linearized quantum gravity. A question that remains is whether a theory independent conclusion may also be drawn by invoking the LOCC--type theorems. The main difficulty in applying these theorems is that they rely on the assumption of the theory possessing a specific tensor decomposition of the state space. In this direction, we suggest that a Bell test would considerably strengthen the case for the importance of detecting GME. Being a theory independent measurement of non-classical behaviour, it does not rely on assumptions about the state space of an underlying theory, but only makes reference to observed data. The violation of a Bell inequality as a result of the experiment allows for a crisp conclusion: \emph{gravitational interactions create  nonlocality}.

Nonlocality is perhaps the most distinctive of quantum phenomena \cite{horodecki2009quantum, brunner2014bell,bell1964einstein,dakic2009quantum}. It is a resource at the basis of quantum information theory, providing quantum advantage over classical computers \cite{nielsen2010quantum}. So it is fitting that the first glimpse of quantum gravity might come from the detection of gravity mediated nonlocality.

\begin{acknowledgments}
\textit{Acknowledgments--} We acknowledge support from the Templeton Foundation, The Quantum Information Structure of Spacetime (QISS) Project (qiss.fr) (the opinions expressed in this publication are those of the author(s) and do not necessarily reflect the views of the John Templeton Foundation)  Grant Agreement No.  61466.
\end{acknowledgments}

\end{document}


\title{Supplementary material: Photonic Implementation of Quantum Gravity Simulator} 

\author{Emanuele Polino}
\author{Beatrice Polacchi}
\author{Davide Poderini}
\author{Iris Agresti}
\author{Gonzalo Carvacho}
\author{Fabio Sciarrino}
\affiliation{Dipartimento di Fisica, Sapienza Universit\`{a} di Roma, P.le Aldo Moro 5, I-00185 Roma, Italy}

\author{Andrea \surname{Di Biagio}}
\affiliation{Dipartimento di Fisica, Sapienza Universit\`a di Roma, Piazzale Aldo Moro 5, Roma, Italy}
\affiliation{Institute for Quantum Optics and Quantum Information (IQOQI) Vienna, Austrian Academy of Sciences, Boltzmanngasse 3, A-1090 Vienna, Austria}

\author{Carlo Rovelli}
\affiliation{Aix-Marseille University, Universit\'e de Toulon, CPT-CNRS, Marseille, France,}
\affiliation{Department of Philosophy and the Rotman Institute of Philosophy, Western University, London ON, Canada,}
\affiliation{ Perimeter Institute, 31 Caroline Street N, Waterloo ON, Canada}

\author{Marios Christodoulou}
\affiliation{Institute for Quantum Optics and Quantum Information (IQOQI) Vienna, Austrian Academy of Sciences, Boltzmanngasse 3, A-1090 Vienna, Austria}
\affiliation{Vienna Center for Quantum Science and Technology (VCQ), Faculty of Physics, University of Vienna, Boltzmanngasse 5, A-1090 Vienna, Austria}

\maketitle


\section{Details on the photonic simulator scheme}
\begin{figure*}[h!]
\includegraphics[width=.99\textwidth]{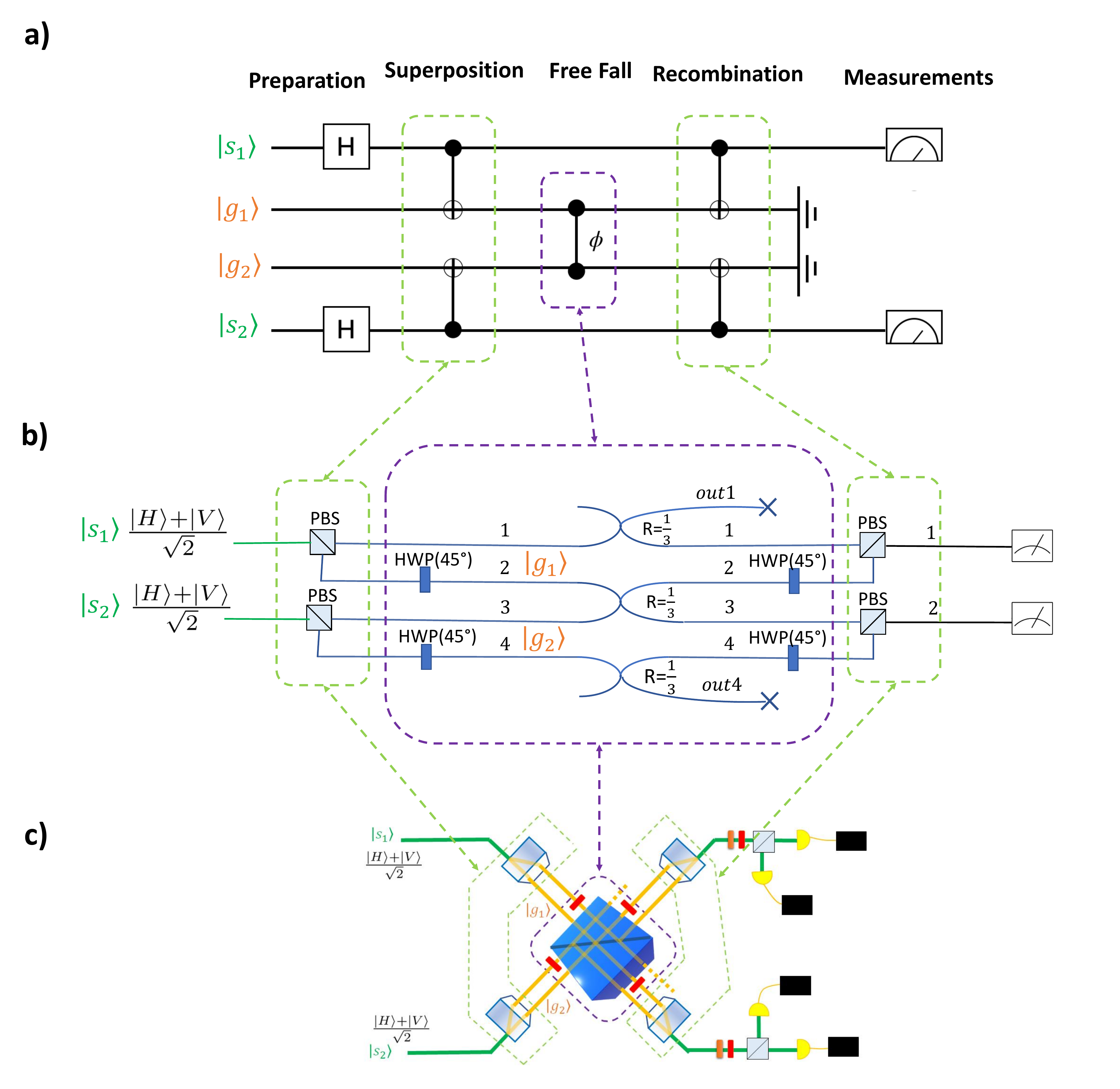}
\caption{\textbf{Linear optical scheme for Quantum Circuit simulator.} Comparison between \textbf{a)} the circuital scheme, \textbf{b)} the optical scheme and \textbf{c)} the realized implementation of the optical scheme. The spin qubits of the simulator are encoded in the polarization degrees of freedom of the two photons, while the geometry degrees of freedom are encoded in their paths. The CZ gate in the path degree of freedom is implemented using the scheme in \cite{obrien2003demonstration}. H=Hadamard gate, PBS=polarizing beam splitter, HWP=half-waveplate.}
\label{fig:sperconf} 
\end{figure*}

The implementation of the Quantum Circuit Simulator (Fig. \ref{fig:sperconf}a) is realized with the linear optical scheme depicted in Fig. \ref{fig:sperconf}b (see Ref.\cite{obrien2003demonstration}) and implemented by the apparatus in Fig. \ref{fig:sperconf}c. We focus now on the linear optical scheme (Fig. \ref{fig:sperconf}b). Here, we map the gravitational field basis states $\{ \ket{g_{LR}}, \ket{g_{LL}}, \ket{g_{RR}}, \ket{g_{RL}} \}$, introduced in Eq. (1) of the main text, to the path degree of freedom of two single photons, represented by the following basis states $\{ \ket{l_1},  \ket{l_2},\ket{l_3}, \ket{l_4} \}$, where the state $\ket{l_i}$ represents a photon along the path $i$   (with $i=1,...,4$). Their spin basis states $\{ \ket{\uparrow},\ket{\downarrow}  \}$, instead, are simulated through the vertical and horizontal photon polarizations $\{ \ket{V},\ket{H}  \}$. Exploiting these two degrees of freedom, each step of the logic circuit can straightforwardly be implemented by means of linear optical elements.

The preparation stage (a Hadamard gate for each spin qubit) of the two polarization states gives rise to the product state $(\ket{H}+\ket{V}))/\sqrt{2} \otimes (\ket{H}+\ket{V}))/\sqrt{2}$. Then, the superposition step is realized by two polarizing beam splitters (PBS) that deterministically perform a control-NOT gate, entangling the polarization and path degrees of freedom of each photon (Fig. \ref{fig:sperconf}b). The global state, tensor product of two entangled states, reads:
\begin{equation}
    \ket{\Psi} = \frac{1}{2} \left[(\ket{H} \ket{l_1}+\ket{V}\ket{l_2})\otimes (\ket{H} \ket{l_3}+\ket{V}\ket{l_4}) \right]\;,
\end{equation}

Then, the path qubits of the two photons, representing the two gravitational fields of the masses, are subject to a control-Phase (CZ) gate,  to mimic gravitational interaction of the masses mediated by gravitational field. In order to act only on the path degree of freedom independently of polarization (according to the assumption that the masses do not interact directly), two half-waveplates rotated by $45^\circ$ are inserted along paths $1$ and $3$. In this way the state becomes:
\begin{equation}
    \ket{\Psi}' = \frac{1}{2} [\,( \ket{l_1}+\ket{l_2})\otimes ( \ket{l_3}+\ket{l_4})\ket{V}\ket{V}\,]\;,
\end{equation}
so that the polarization and path are factorized. This allows the ``gravitational gate" (Free Fall stage in Fig.\ref{fig:sperconf}a) to act only on the path degree of freedom.

At this point, it is convenient to switch to the second quantization formalism describing photon number states along the modes through annihilation and creation operators.

In this formalism the state of a photon along a mode $\bm{k}$, that comprises all photon's degrees of freedom, can be described by the annihilation and creation operators, denoted by $a_{\bm{k}}$ and $ a^\dagger_{\bm{k}}$, respectively. 
Such operators obey the following bosonic commutation rules: 
\begin{equation}\label{commann}
    [a_{\bm{k_i}},a_{\bm{k_j}}]=[a_{\bm{k_i}}^\dagger,a_{\bm{k_j}}^\dagger]=0 \qquad \qquad 
[a_{\bm{k_i}},a_{\bm{k_j}}^\dagger]=\delta_{ij}\;,
\end{equation}
where $\bm{k_i}$ and $\bm{k_j}$ are two modes of the field.

Fock states, represented by $\ket{n_{\bm{k}}, \ldots}$, are states with a fixed number $n_{\bm{k}}$ of photons along each mode. The action of annihilation (creation) operators on such states is to destroy (create) a photon along mode $\bm{k}$, according to the relations:
\begin{equation}
    a_{\bm{k}}\ket{n_{\bm{k}}}= \sqrt{n_{\bm{k}}}\ket{n_{\bm{k}}-1}\qquad 
 a_{\bm{k}}^\dagger\ket{n_{\bm{k}}}= \sqrt{n_{\bm{k}}+1}\ket{n_{\bm{k}}+1}\;.
\end{equation}

In particular, one can generate any Fock state from  vacuum state $|0\rangle$, by iteratively applying creation operators on the modes:
$|n_{\bm{k}}\rangle=\frac{a^{\dagger\, n_{\bm{k}}}_{\bm{k}}}{\sqrt{n_{\bm{k}}!}}|0\rangle$.

The unitary evolution $U$ on operators $a_i^{\dagger}$ can be described by the following relation:
\begin{equation}\label{opev}
 a_i^{\dagger} \rightarrow \sum_{j}U_{ij}^{\dagger}\,b_j^{\dagger}\;,  
\end{equation}
where $b_j^{\dagger}$ represent the output modes of the transformation.

We now describe the scheme \cite{obrien2003demonstration} which performs the control-Phase in the path degree of freedom of photons in our platform. Since the polarization state factorizes, it is omitted.

Such a scheme is composed of three parallel two-mode couplers, that  allow the interference of two input modes. The expression of each coupler is: 
\begin{equation} \label{eq:couplermatrix}
U_{R} = \left( \begin{array}{cccc}
i\sqrt{R} & \sqrt{1-R}  \\
\sqrt{1-R} & i \sqrt{R}   
\end{array} \right)\; ,
\end{equation}
For instance, consider the upper coupler between modes $1$ and $out1$ in Fig. \ref{fig:sperconf}b. The mode corresponding to the operator $a_1^{\dagger}$ can be transmitted with probability amplitude  $\sqrt{1-R}$ and reflected with amplitude $i\sqrt{R}$. Therefore it evolves as: $a_1^{\dagger} \rightarrow  \sqrt{1-R}\,\, b_{out1}^{\dagger}-i\sqrt{R} \,\, b_{1} ^{\dagger}$, where $b_{1}^{\dagger}$ and $b_{out1}^{\dagger}$ represent the transmitted and reflected creation mode operators, respectively.
Three parallel couplers  $U_{out1\,,1}$, $U_{2\,,3}$, $U_{4\,,out4}$, with $R=1/3$, allow to interfere the mode pairs ($out1,1$), ($2,3$) and ($4,out4$), respectively. Modes $out1$ and $out4$ are vacuum ancillary modes whose utility will be clarified later. Consider all four possible two-photon states in which we have a photon along mode $1$ or $2$ (first photon) and a photon along mode $2$ or $4$ (second photon).

Exploiting relation \eqref{opev} and definition \eqref{eq:couplermatrix}, one obtains, for each considered input state, the following transformation under the global evolution $U_{CZ}=U_{out1\,,1}\otimes U_{2\,,3}\otimes U_{4\,,out4}$:

\begin{equation}\begin{split}
   &\ket{l_1}\ket{l_3} \equiv | 1\rangle_{1}| 1\rangle_{3}=a_1^{\dagger}a_3^{\dagger}| 0\rangle \xrightarrow{U_{CZ}}\\ &\left(\sqrt{\frac{2}{3}}b_{out1}^{\dagger}-i\frac{1}{\sqrt{3}}b_{1}^{\dagger}\right)\, \left(\sqrt{\frac{2}{3}}b_{2}^{\dagger}-i\frac{1}{\sqrt{3}}b_3^{\dagger}\right)\,| 0\rangle= 
\\
&=\frac{2}{3}| 1\rangle_{out1}| 1\rangle_{2}-i\frac{\sqrt{2}}{3} (| 1\rangle_{out1}| 1\rangle_{3} +| 1\rangle_{1}| 1\rangle_{2})-\frac{1}{3}| 1\rangle_{1}| 1\rangle_{3}\;,
\end{split}
\end{equation}

\begin{equation}\begin{split}
   &\ket{l_1}\ket{l_4} \equiv| 1\rangle_{1}| 1\rangle_{4}=a_1^{\dagger}a_4^{\dagger}| 0\rangle \xrightarrow{U_{CZ}}\\ &\left(\sqrt{\frac{2}{3}}b_{out1}^{\dagger}-i\frac{1}{\sqrt{3}}b_{1}^{\dagger}\right)\, \left(\sqrt{\frac{2}{3}}b_{out4}^{\dagger}-i\frac{1}{\sqrt{3}}b_4^{\dagger}\right)| 0\rangle= 
\\
&=\frac{2}{3}| 1\rangle_{out1}| 1\rangle_{out4}-i\frac{\sqrt{2}}{3} (| 1\rangle_{out1}| 1\rangle_{4} +| 1\rangle_{1}| 1\rangle_{out4})+\\
&-\frac{1}{3}| 1\rangle_{1}| 1\rangle_{4}\;,
\end{split}
\end{equation}

\begin{equation}\begin{split} \label{eq:czinterfterm}
   &\ket{l_2}\ket{l_3} \equiv| 1\rangle_{2}| 1\rangle_{3}=a_2^{\dagger}a_3^{\dagger}| 0\rangle \xrightarrow{U_{CZ}}\\ &\left(\sqrt{\frac{2}{3}}b_{3}^{\dagger}-i\frac{1}{\sqrt{3}}b_2^{\dagger}\right)\, \left(\sqrt{\frac{2}{3} } b_2^{\dagger}-i \frac{1}{\sqrt{3}} b_3^{\dagger}\right)| 0\rangle= 
\\
&=\frac{2}{3}| 1\rangle_{3}| 1\rangle_{2}-i\frac{2 \sqrt{2}}{3} (| 2\rangle_{3}+| 2\rangle_{2})-\frac{1}{3}| 1\rangle_{2}| 1\rangle_{3}= 
\\
&=\frac{1}{3}| 1\rangle_{2}| 1\rangle_{3}-i\frac{2\sqrt{2}}{3} (| 2\rangle_{3}+| 2\rangle_{2})\;,
\end{split}
\end{equation}

\begin{equation}\begin{split}
   &\ket{l_2}\ket{l_4} \equiv | 1\rangle_{2}| 1\rangle_{4}=a_2^{\dagger}a_4^{\dagger}| 0\rangle \xrightarrow{U_{CZ}}\\ &\left(\sqrt{\frac{2}{3}}b_{3}^{\dagger}-i\frac{1}{\sqrt{2}}b_2^{\dagger}\right)\, \left(\sqrt{\frac{2}{3}}b_{out4}^{\dagger}-i\frac{1}{\sqrt{3}}b_4^{\dagger}\right)\,| 0\rangle= 
\\
&=\frac{2}{3}| 1\rangle_{3}| 1\rangle_{out4}-i\frac{\sqrt{2}}{3} (| 1\rangle_{3}| 1\rangle_{4} +| 1\rangle_{2}| 1\rangle_{out4})-\frac{1}{3}| 1\rangle_{2}| 1\rangle_{4}\;,
\end{split}
\end{equation}
where $\ket{n}_x$ indicates the Fock state with $n$ photons along $x$ mode and  the last equality of Eq. \eqref{eq:czinterfterm} used the indistinguishability of the photons. Post-selecting  the final states where there are one and only one photon along modes $1$ or $2$, and simultaneously one and only one photon along modes $3$ or $4$, the above transformations reduce to the following four transformations: $\ket{1}_1\,\ket{1}_3\xrightarrow{U_{CZ}}-\ket{1}_1 \ket{1}_3/3$ , $\ket{1}_1 \ket{1}_4\xrightarrow{U_{CZ}}-\ket{1}_1 \ket{1}_4/3$ , $\ket{1}_2 \ket{1}_3\xrightarrow{U_{CZ}}+\ket{1}_2 \ket{1}_3/3$ , $\ket{1}_2 \ket{1}_4\xrightarrow{U_{CZ}}-\ket{1}_2 \ket{1}_4/3$ . 
Each term evolves to a post-selected state with probability $1/9$ . Hence,  multiplying each term to $-1$ and defining for simplicity the logic states $\ket{0}_T\equiv \ket{l_1}=\ket{1}_1$ , $\ket{1}_T\equiv \ket{l_2}=\ket{1}_2$ , $\ket{0}_C\equiv \ket{l_4}=\ket{1}_4$ , $\ket{1}_C\equiv \ket{l_3}=\ket{1}_3$,  the post-selected transformation is: 

\begin{equation}
\begin{cases}\left| 0\right\rangle_T |0\rangle_C \phantom{-}\longrightarrow \phantom{-} \left| 0\right\rangle_T |0\rangle_C \\
\left| 0\right\rangle_T |1\rangle_C \phantom{-}\longrightarrow  \phantom{-}\left| 0\right\rangle_T |1\rangle_C \\
\left| 1\right\rangle_T |0\rangle_C \phantom{-}\longrightarrow  \phantom{-}\left| 1\right\rangle_T |0\rangle_C \\
\left| 1\right\rangle_T |1\rangle_C \phantom{-}\longrightarrow  -\left| 1\right\rangle_T |1\rangle_C \\
\end{cases}\;,
\end{equation}
that corresponds to a control-Phase operation between the two qubits encoded in the path degree of freedom of the photons.

Such a scheme is implemented by the setup in Fig. \ref{fig:sperconf}c where a central beam splitter acts on different modes interfering in three points, while four beam displacers act pairwise to split and recombine the photons based on their polarizations.

\section{Experimental details}

To quantify the degree of indistinguishability of the photons imprinting the BS, we measured the visibility of the Hong-Ou-Mandel (HOM) dip of the coincidences with respect to the time delay between the two photons. The experimental value found for the visibility is $V^{exp}=0.73 \pm 0.02$ that we compare to the ideal (perfect indistinguishable photons) theoretical one $V^{theo}=0.8$, obtaining a ratio equal to $V^{exp}/V^{theo}=0.913 \pm 0.025$.

The calcite beam displacers act as the entangling gates of the Superposition and Recombination stages between path and polarization of the single photons with a fidelity $>99.5\%$.

The measured value of the reflectivities of the beam splitter is $|r_H|^2=0.329\pm0.001$ for the horizontal polarization and $|r_V|^2=0.337\pm0.001$ for the vertical polarization of the incoming photons.

The fidelity of the scheme is also affected by the degree of indistinguishability of the interfering photons in all their degrees of freedom. Polarization, frequency, time of arrival and spatial mode overlap all affect indistinguishability. Time of arrival and spatial mode overlap are crucial: the arrival time on the BS is controlled by suitable delay lines, while spatial modes are recombined by fine alignment through optical mirrors. 

\section{Entanglement degrading effects and quantum state tomographies}
In the experimental simulation, we studied two kinds of noisy  effects that could be observed in future massive experiments, preventing the formation or the detection of entanglement.

\begin{figure}[h]
\includegraphics[width=.85\textwidth]{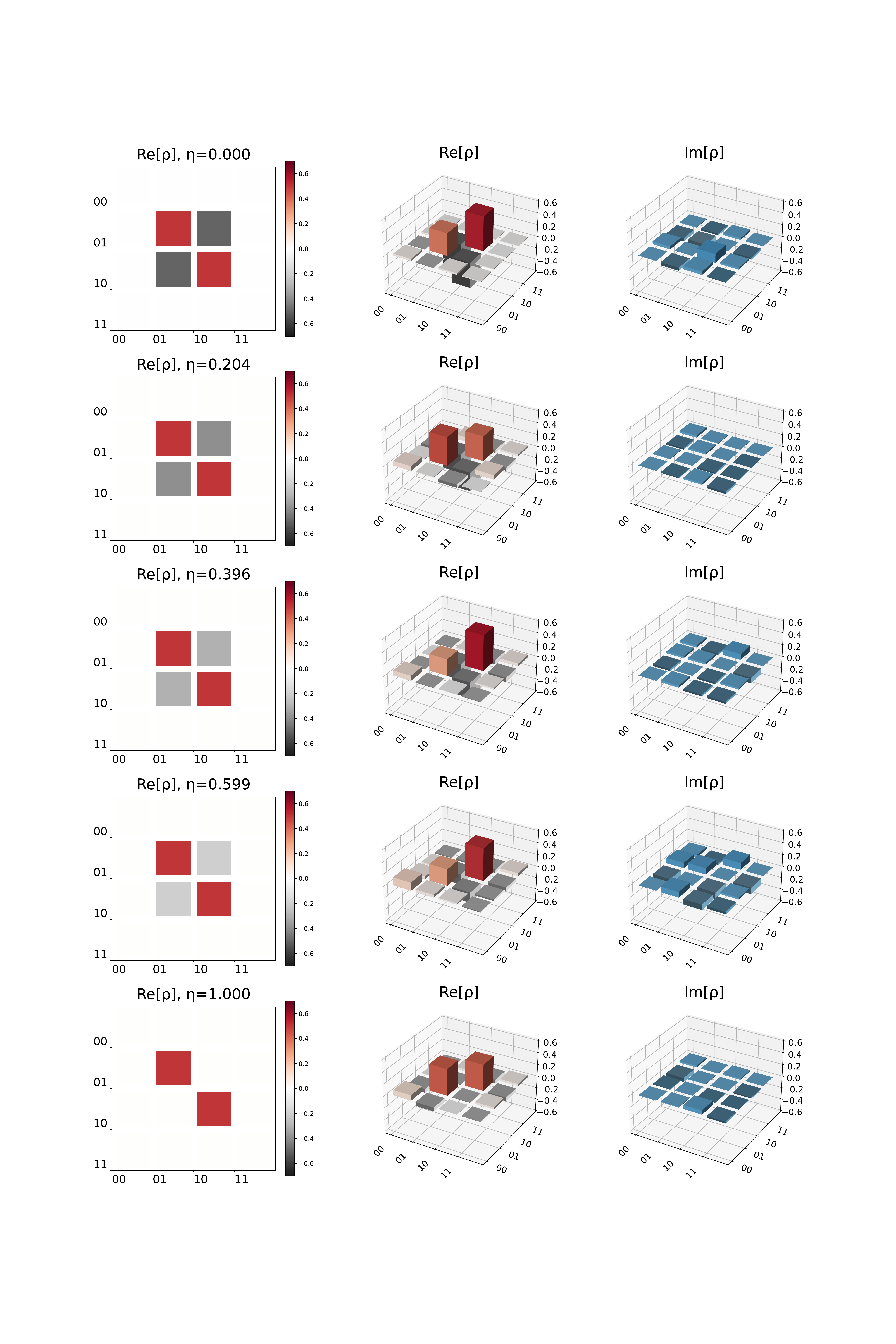}
\caption{\textbf{Results of QC simulator with decoherence effect.} 
 Experimental real (left) and imaginary (right) parts of the measured density matrices of the polarization  states of the spin qubits, generated after the Free Fall stage and post selection of the geometry qubits, as function of the degree of decoherence $\eta$ in Eq.(11) of the main text.    
}\label{fig:expres}
\end{figure}

The first effect studied, as explained in the main text, is the decoherence of the state that could be due to either gravitational induced collapse or to the coupling of the degrees of freedom of the masses with the environment and is one of the major obstacles to a proper realization of massive GME experiments. In Fig. \ref{fig:expres}, we show the quantum state tomographies of states with different degrees of coherence, i.e. for different parameter values $\eta$  in Eq. (11) of the main text. Decoherence was induced by using birefringent plates of different widths. Birefringence causes horizontal and vertical polarizations to experience different refraction indices, and they are thus time delayed one with respect to the other, with a time-delay depending on the plate width.

The second effect studied is the temporal indistinguishability of the photon in the degree of freedom of time arrival: if the two photons run across the interferometer at different times, they will not interfere and no entanglement will arise between them. This can simulate the effect of non-synchronized time of the masses, or a screened interaction that would prevent the generation of entanglement.
In Fig.\ref{fig:expres} we show the quantum state tomographies of states with different degrees of distinguishability $1-v$ in Eq.(12) of the main text, which is tuned by varying the delay between the two photons before the beam splitter.

\begin{figure}[h]
\includegraphics[width=.85\textwidth]{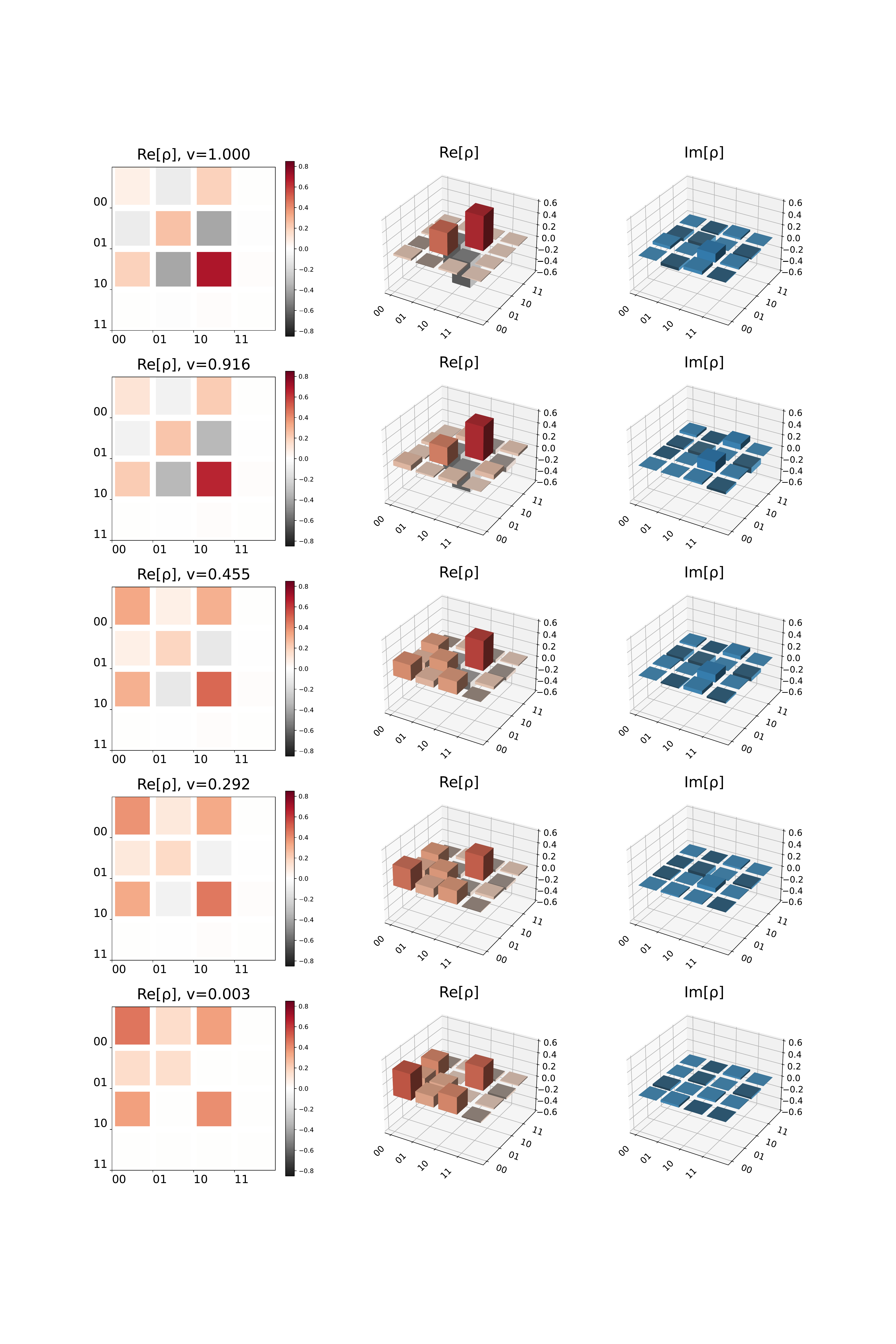}
\caption{\textbf{Results of QC simulator with varying photon distinguishability.} 
 Experimental real (left) and imaginary (right) parts of the measured density matrices of the polarization states of the spin qubits, generated after the Free Fall stage and post selection of the geometry qubits, at different temporal delays between the two photons. The distinguishability degree $\left(1-v \right)$ extracted from fits on the experimental data is reported above each tomography.
}\label{fig:expres_hom}
\end{figure}

\begin{figure}[h]
\includegraphics[width=.7\textwidth]{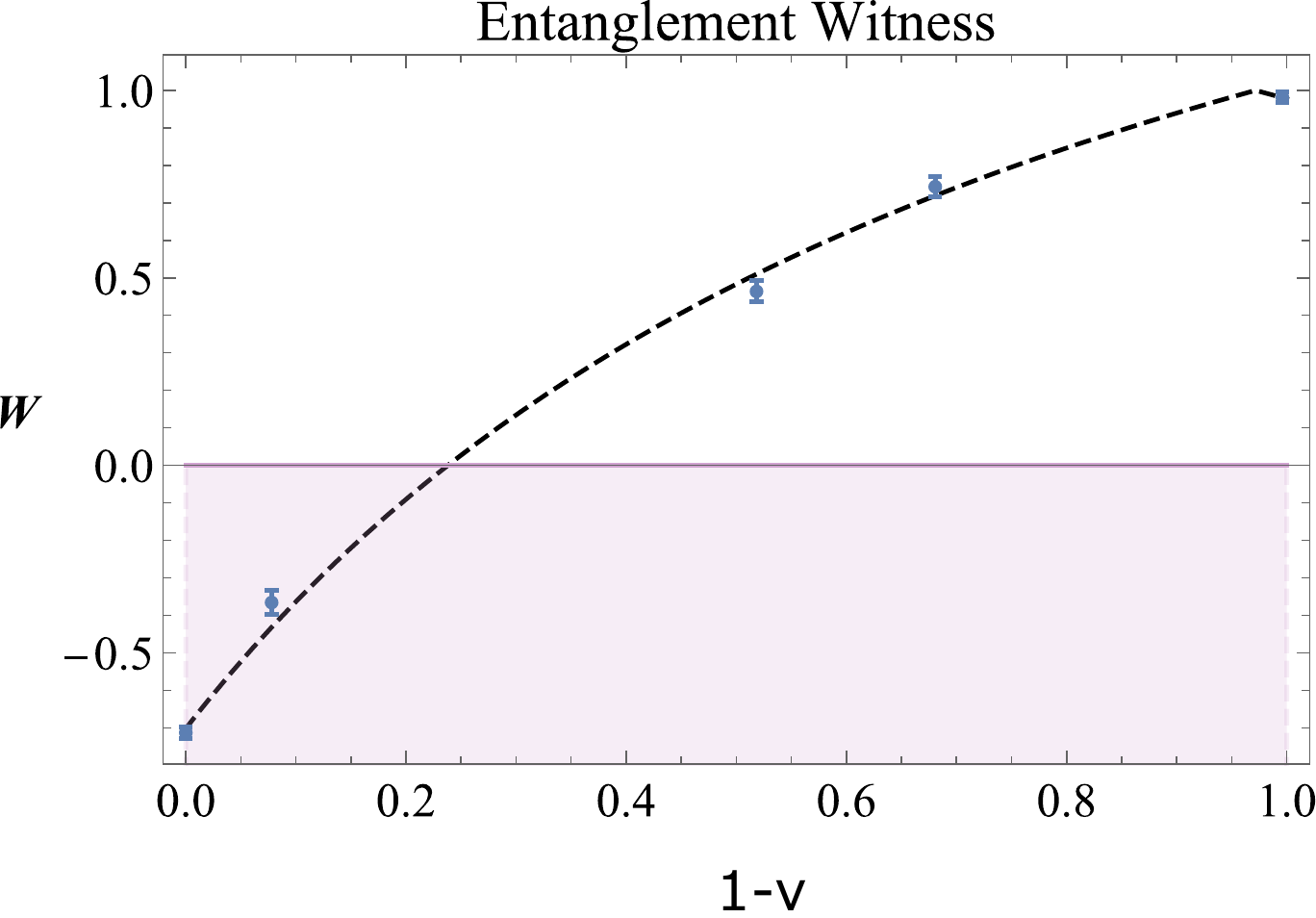}
\caption{\textbf{Results of QC simulator with delays between the time arrivals of the photons.} 
 Values of the witness $\mathcal{W}$ measured as function of the degree of distinguishability $\left( 1-v \right)$ that is varied by changing the relative time delay between the photons.  All error bars are due to Poissonian statistics of the measured events. The purple line indicates the value, above which the state does not violate the entanglement witness, and the shadowed area indicates the region where the witness certifies the entanglement of the state.  The dashed black line represents the theoretical curve from the model of the experimental setup.  
}\label{fig:delaywit}
\end{figure}

Finally, for the different degrees of indistinguishability we also measured the entanglement witness as reported in Fig.\ref{fig:delaywit}.

